\documentclass[%
 reprint,%
 amssymb, amsmath,%
 aip,apl,%
]{revtex4-1}

\usepackage{graphicx}%
\usepackage{dcolumn}%
\usepackage{bm}%

\begin{document}

\title{Dynamic Calibration of Higher Eigenmode Parameters of a Cantilever in Atomic Force Microscopy Using Tip-Surface Interactions}

\author{Stanislav S. Borysov}
\email{borysov@kth.se}

\affiliation{Nanostructure Physics, KTH Royal Institute of Technology, Roslagstullsbacken 21, SE-106 91 Stockholm, Sweden}
\affiliation{Nordita, KTH Royal Institute of Technology and Stockholm University, Roslagstullsbacken 23, SE-106 91 Stockholm, Sweden}
\affiliation{Theoretical Division, Los Alamos National Laboratory, Los Alamos, NM 87545, USA}

\author{Daniel Forchheimer}
\affiliation{Nanostructure Physics, KTH Royal Institute of Technology, Roslagstullsbacken 21, SE-106 91 Stockholm, Sweden}

%\author{Alexander V. Balatsky}
%\affiliation{Nordita, KTH Royal Institute of Technology and Stockholm University, Roslagstullsbacken 23, SE-106 91 Stockholm, Sweden}
%\affiliation{Institute for Materials Science, Los Alamos National Laboratory, Los Alamos, NM 87545, USA}

\author{David B. Haviland}
\affiliation{Nanostructure Physics, KTH Royal Institute of Technology, Roslagstullsbacken 21, SE-106 91 Stockholm, Sweden}
\email{haviland@kth.se}

\date{\today}%
%\revised{}%

\begin{abstract}
We present a theoretical framework for the dynamic calibration of the higher eigenmode parameters (stiffness and optical lever responsivity) of a cantilever. The method is based on the tip-surface force reconstruction technique and does not require any prior knowledge of the eigenmode shape or the particular form of the tip-surface interaction. The calibration method proposed requires a single-point force measurement using a multimodal drive and its accuracy is independent of the unknown physical amplitude of a higher eigenmode.
\end{abstract}

\maketitle

%-------------------------------------------------------------------------------
{\it Introduction.}---Atomic force microscopy \cite{AFM1} (AFM) is one of the primary methods of the surface analysis, reaching resolution of nanometers and below. In a conventional AFM an object is scanned using a microcantilever with a sharp tip at the free end. Measuring cantilever deflections allows not only for the reconstruction of the surface topography but also provides insight into various material properties\cite{forceIterpret1,forceIterpret2}. If cantilever deflection is measured near one of its resonance frequencies, an enhanced force sensitivity is achieved due to multiplication by the sharply peaked cantilever transfer function. Measurement of response at multiple eigenmodes can provide additional information about the tip-surface interactions \cite{multi1, multi10, multi2, multi3, multi4, multi5, multi6, multi7}.

\begin{figure}
\centering
\includegraphics[width=80mm,angle=0]{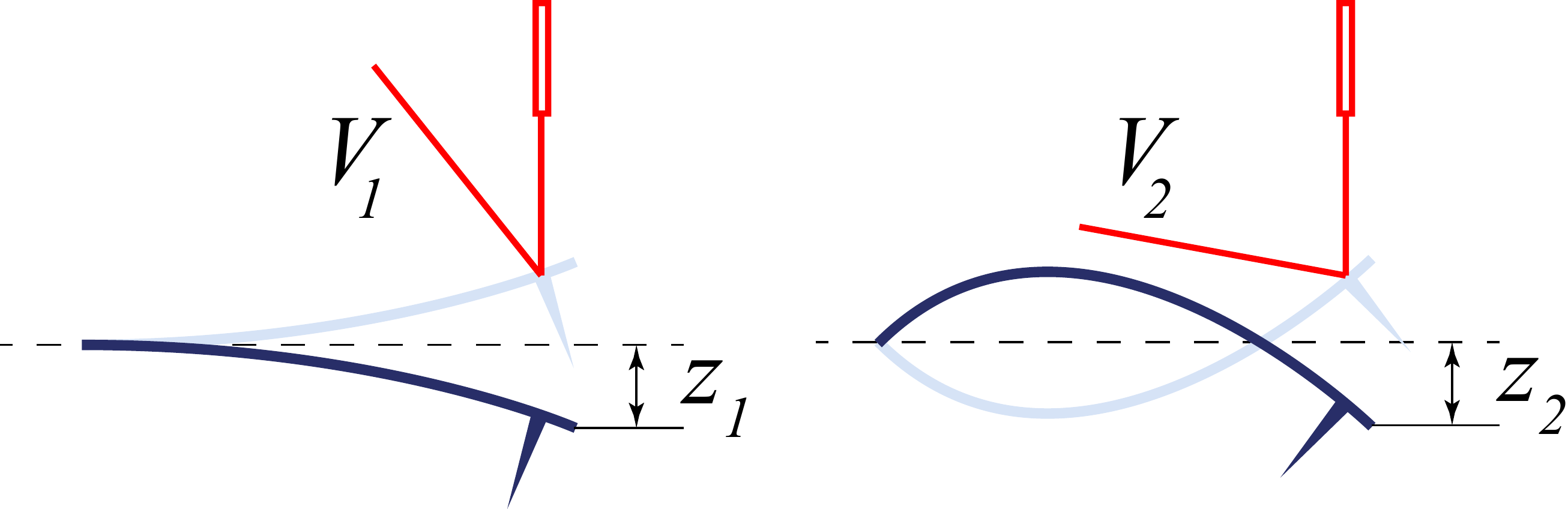}
 \caption{(Color online) Schematic illustration of the two first flexural eigenmode shapes of a rectangular cantilever and an optical detection system. Measuring of the slope at the free end leads to the situation when the equal vertical tip deflections, $z_1=z_2$, result in the different detected voltages, $V_1\neq V_2$. In the case of small deflections, $z_n\propto V_n$ with some coefficient $\alpha_n$ called optical lever responsivity.}
\label{fig:cant}
\end{figure}

The optical detection system\cite{meyer:2089} common to most of AFM systems leverages a laser beam reflected from the cantilever, measuring the slope rather than its vertical deflection. This underlying principle leads to the measured voltage at the detector being dependent on the geometric shape of the excited eigenmode (Fig.~\ref{fig:cant}). While determination of the stiffness and optical lever resposivity\cite{Note1} of the first flexural eigenmode can be performed with high accuracy using a few well-developed techniques \cite{calib4,calib5,calib6,calib9,calib10,calib7, calib13, calib12, calib15}, calibration of the higher eigenmode parameters is still a challenging task. The main problem with the existing theoretical approaches based on the calculation of eigenmode shapes is that real cantilevers differ form the underlying solid body mechanical models due to the tip mass\cite{calib11}, fabrication inhomogeneities and defects\cite{calib8}, etc.
In this letter, we propose a method which overcomes these deficiencies. 

The method uses the fact that the tip-surface force is equally applied to all eigenmodes \cite{Note2} Any other force acting on the whole cantilever, e.g. of thermal or electromagnetic nature, should be convoluted with the eigenmode shape, leading to a different definition of the effective dynamic stiffness. Thus, knowledge of the cantilever's geometry is not required to reconstruct tip-surface force. The framework proposed harnesses a force reconstruction technique inspired by the Intermodulation AFM \cite{imafm1} (ImAFM) \cite{}, which was recently generalized to the multimodal case\cite{PhysRevB.88.115405}. It is worth noting that the proposed calibration method is similar to that described in Ref.~\onlinecite{sugimoto:093120}, where stiffness of the second eigenmode is experimentally defined using consecutive measurements of the frequency shift caused by the tip-surface interaction for different eigenmodes. In contrast, we propose a simultaneous one-point measurement using a multimodal drive which avoids issues related to the thermal drift\cite{1650470} and it exploits nonlinearities for higher calibration precision.

%-------------------------------------------------------------------------------
{\it Cantilever model.}---We consider a point-mass approximation of a cantilever derived from the eigenmode decomposition of its continuum mechanical model, e.g. the Euler-Bernoulli beam theory. 
Such a reduced system of coupled harmonic oscillators in the Fourier domain has the following form
\begin{equation}
\label{eq:sys}
k_n\alpha_n\hat V_n(\omega) = \hat G_n(\omega)\left[\hat F(\omega) + \hat{\mathrm{f}}_n(\omega)\right],
\end{equation}
where the hat denotes the Fourier transform, $\omega$ is the frequency, $k_{n}$ is the effective dynamic stiffness of the $n$th eigenmode ($n=1,\dots,N$), $\alpha_{n}$ is the optical lever responsivity, $V_n$ is the measured voltage (corresponding to the eigencoordinate $z_{n} = \alpha_{n}V_{n}$, where total tip deflection is $z=\sum_{n=1}^nz_n$), $\hat G_{n} = [1 + (i/Q_{n})(\omega/\omega_{n}) - (\omega/\omega_{n})^2]^{-1}$ is the linear transfer function of a harmonic oscillator with the resonant frequency $\omega_{n}$ and quality factor $Q_{n}$, $F$ is a nonlinear tip-surface force and $\mathrm{f}_n$ is a drive force. The stiffness is deliberately excluded from the expression for the $G_n$ since the parameters $Q_n$ and $\omega_n$ can be found employing the thermal calibration method \cite{calib5,calib10}. Note that if the force amplitudes on the right hand side of Eq.~(\ref{eq:sys}) are known, one immediately gets $k_n$ and $\alpha_n$ taking the absolute values in combination with the equipartition theorem 
\begin{equation}
\label{eq:equipart}
	k_{n}\left\langle z_n^2\right\rangle = k_{n}\alpha_{n}^2\left\langle V_{n}^2\right\rangle = k_BT,
\end{equation}
where $\langle\cdot\rangle$ is a statistical average, $k_B$ is the Boltzmann constant and $T$ is an equilibrium temperature. 

%-------------------------------------------------------------------------------
{\it Spectral fitting method.}---The task at hand requires reconstruction of the forces on the right hand side of Eq.~(\ref{eq:sys}) from the measured motion specrtum. Firstly, it is possible to remove the unknown drive contribution, $\hat{\mathrm{f}}_n$, for each $n$, by means of subtraction of the free oscillations spectrum, $\hat V_{n}^\mathrm{f}$ (far from the surface, where $F\equiv 0$), from the spectrum of the engaged tip motion, $\hat V_{n}^\mathrm{e}$ (near the surface). It gives the following relationships
\begin{equation}
\label{eq:sys2}
k_n\alpha_n\Delta\hat V_n = \hat G_n\hat F,
\end{equation}
where $\Delta \hat V_{n} \equiv \hat V_{n}^\mathrm{e} - \hat V_{n}^\mathrm{f}$. 
For the high-$Q$ cantilevers, the measured response near resonances allows to detect each $\hat V_{n}$ separately with the high signal-to-noise ratio (SNR). Neglecting possible surface memory effects, $F$ depends on the tip position $z$ and its velocity $\dot z$ only. With this assumption, the force model to be reconstructed has some generic form
\begin{equation}
\label{eq:force_model}
\begin{array}{lcl}
	\tilde F(z,\dot z) &=& \sum\limits_{i=0}^{P_z}\sum\limits_{j=0}^{P_{\dot z}}g_{ij} z^i\dot z^j \\ &=& \sum\limits_{i=0}^{P_z}\sum\limits_{j=0}^{P_{\dot z}}g_{ij} \left(\sum\limits_{n=1}^{N}\alpha_nV_n^\mathrm{e}\right)^i  \left(\sum\limits_{n=1}^{N}\alpha_n\dot V_n^\mathrm{e}\right)^j,
\end{array}
\end{equation}
with $P=P_zP_{\dot z}-1$ unknown parameters $g_{ij}$ ($g_{00}$ is excluded because it corresponds to the static force) which can be found using the spectral fitting method \cite{imafm2,imafm3}: Substitution of Eq.~(\ref{eq:force_model}) in Eq.~(\ref{eq:sys2}) yields a system of linear equations for $g_{ij}$. However, this system becomes nonlinear with respect to unknown $k_n$ and $\alpha_n$. 

%-------------------------------------------------------------------------------
{\it Intermodulation AFM.}---Assuming that $\alpha_1$ and $k_1$ are calibrated using one of the methods mentioned in the Introduction, the resulting system contains $2(N-1) + P$ unknown variables. Use of the equipartition theorem [Eq.~(\ref{eq:equipart})] for each eigenmode gives us $N-1$ equations and the remaining equations should be defined using Eq.~(\ref{eq:sys2}) for the known response components in the motion spectrum. If the force acting on a tip over its motion domain is approximately linear ($P=1$), one drive tone at each resonant frequency is enough to determine the system. However, when force behaves in a nonlinear way ($P>1$), as is usually the case, more measurable response components in the frequency domain are needed. The core idea of ImAFM relies on the ability of a nonlinear force to create intermodulation of discrete drive tones in a frequency comb. Driving an eigenmode subject to a nonlinear force on at least two frequencies $\omega_{n1}^\mathrm{d}$ and $\omega_{n2}^\mathrm{d}$, gives response in the frequency domain at these drive frequencies and their higher harmonics but also at their linear combinations $n\omega_{1}^\mathrm{d}+m\omega_{2}^\mathrm{d}$ ($n$ and $m$ are integers) called intermodulation products (IMPs). Use of the small base frequency $\delta\omega = |\omega_{n1}^\mathrm{d} - \omega_{n2}^\mathrm{d}|$ results in concentration of IMPs close to the resonance which opens the possibility for their detection with high SNR. This additional information can be used in Eq.~\ref{eq:sys2} for the reconstruction of nonlinear conservative and dissipative forces \cite{imafm2,imafm3,PhysRevB.88.115405,imafm6} with the only restriction that IMPs in the different narrow bands near resonances contain the same information about the unknown force parameters \cite{PhysRevB.88.115405}. 

\begin{figure}
\centering
\includegraphics[width=75mm,angle=0]{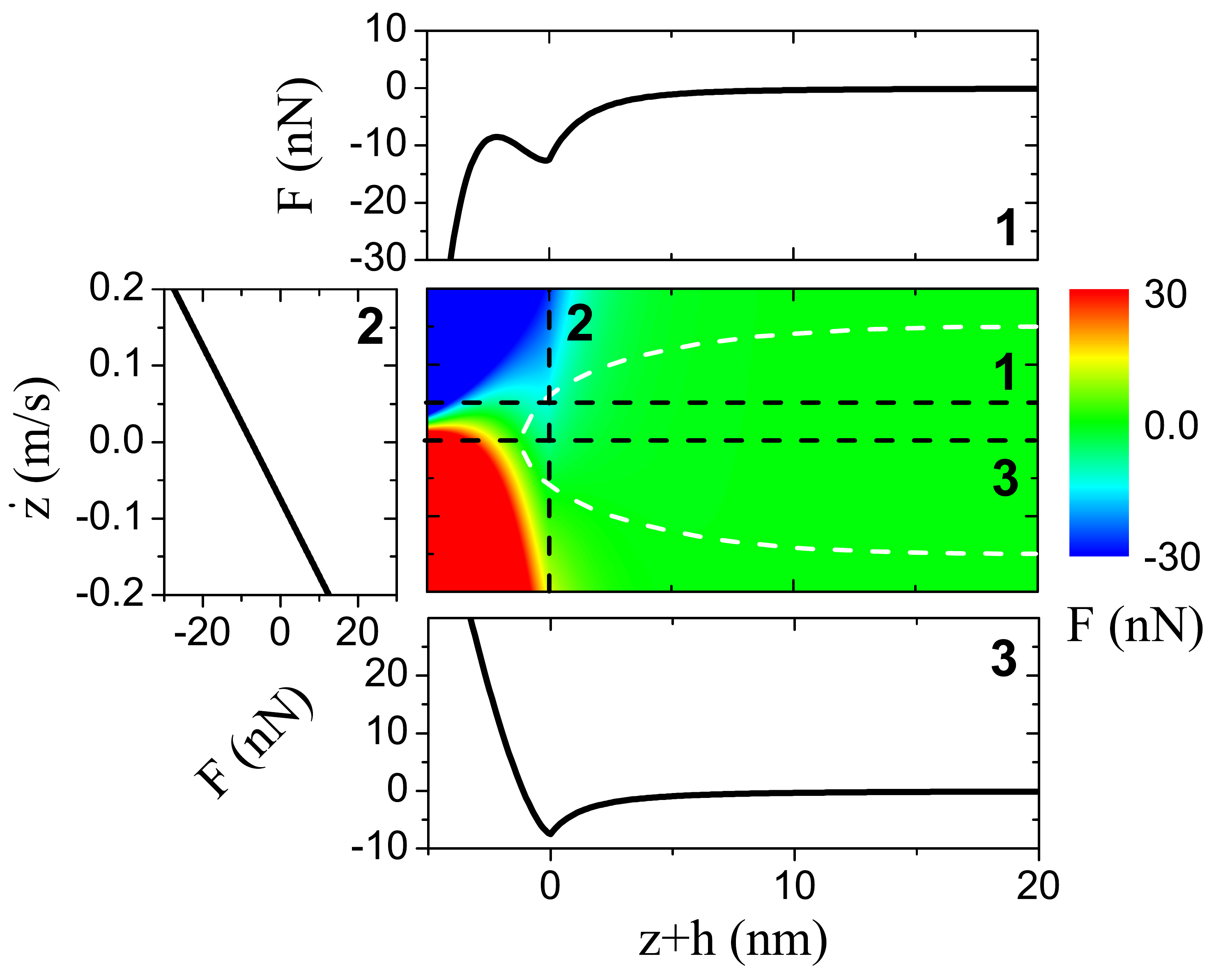}
 \caption{(Color online) The tip-surface force [Eq.~(\ref{eq:vdW-DMT})] used in the simulations. White dashed line corresponds to a phase space trajectory of the bimodal stiff cantilever with the eigenmode amplitudes $A_1=A_2=12.5$~nm and reference height $h=17$~nm. Cross-sections for different values of $z$ and $\dot z$ are shown: The projections (1) and (2) correspond to the lines $\dot z = 0.05$ m/s and $z = 0$ nm respectively; the conservative part (3) corresponds to the line $\dot z = 0$ m/s.}
\label{fig:actual_force}
\end{figure}

%-------------------------------------------------------------------------------
{\it Calculation details.}---In the rest of the paper, we consider a bimodal case implying straightforward generalization for $N>2$ eigenmodes. Equation~\ref{eq:sys} is integrated using CVODE\cite{CVODE} for two different sets of cantilever parameters from Table \ref{tab:cantilevers}. The cantilever is excited using multifrequency drive (specified below) with frequencies being integer multiples of the base frequency $\delta\omega=2\pi\,0.1$~kHz. The tip-surface force $F$ is represented by the vdW-DMT model\cite{vdW-DMT} with the nonlinear damping term being exponentially dependent on the tip position\cite{PhysRevB.60.11051} 
\begin{equation}
\label{eq:vdW-DMT}
\begin{array}{lcl}
	F = F^{\mathrm{con}} + F^{\mathrm{dis}},\\
	F^{\mathrm{con}}(z) = \begin{cases} 
		-\frac{HR}{6(z+h)^2}, &z+h\geq a_0 \\
		-\frac{HR}{6a_0^2}+\frac{4}{3}E^{*}\sqrt{R(a_0-(z+h))}, &z+h<a_0
	\end{cases}\\
	F^{\mathrm{dis}}(z,\dot z) = - \gamma_1\dot ze^{-(z+h)/\lambda_z},
\end{array}
\end{equation}
where $h$ is a reference height. Its conservative part, $F^{\mathrm{con}}$, has four phenomenological parameters: the intermolecular distance $a_0=0.3$~nm, the Hamaker constant $H=7.1\times 10^{-20}$~J, the effective modulus $E^{*}=1.0$~GPa and the tip radius $R=10$~nm. The dissipative part, $F^{\mathrm{dis}}$, depends on the damping factor $\gamma_1=2.2\times 10^{-7}$ kg/s and the damping decay length $\lambda_z=1.5$~nm. The force [Eq.~(\ref{eq:vdW-DMT})] and its cross-sections are depicted in Fig.~\ref{fig:actual_force}.

\begin{table*}
\caption{\label{tab:cantilevers}Cantilever parameters used for the numerical calculations in the paper. Last column $E$ is a total oscillation energy of a free cantilever with the equal eigenmode amplitudes $A_1=A_2=1$~nm.}
\begin{ruledtabular}
\begin{tabular}{lcccccccc}
Cantilever & $\omega_1$ $(2\pi)^{-1}$KHz & $\omega_2/\omega_1$ & $Q_1$ & $Q_2/Q_1$ & $k_1$~N/m & $k_2/k_1$ & $\alpha_2/\alpha_1$ & $E$ (fJ) \\
1. Soft & 82.7 & 6.35 & 220.0 & 2.9 & 5.0 & 40.0 & 2.0 & 1.02 \\
2. Stiff & 300.0 & 6.3 & 400.0 & 3.0 & 40.0 & 50.0 & 2.0 & 0.105 \\
\end{tabular}
\end{ruledtabular}
\end{table*}

%-------------------------------------------------------------------------------
{\it Calibration using a nonlinear tip-surface force.}---In order to find $k_2$ and $\alpha_2$ from the nonlinear system [Eq.~(\ref{eq:equipart}) and Eq.~(\ref{eq:sys2})], we first solve the linear system for the force parameters $g_{ij}$ . It is then convenient to compare only the conservative part of the tip-surface force given its nonmonotonic behavior. There are two methods to require equality of the reconstructed forces $\tilde F^{(1)}$ (using the band near the first eigenmode) and $\tilde F^{(2)}$ (near the second eigenmode). The first method is to check the difference between the corresponding parameters $g_{ij}^{(1)}$ and $g_{ij}^{(2)}$. However, this approach is not suitable because two completely different sets of coefficients might define very similar functions on the interval of the actual engaged tip motion, $[A^\mathrm{min,e}; A^\mathrm{max,e}]$, where $A^\mathrm{max} = \max A(t) = \max z(t)$. As numerical simulations have shown, the error function does not have a well-defined global minimum and it is highly sensitive to reconstruction errors. An alternative approach is to minimize a mean square error function in the real space
\begin{equation}
\label{eq:error_ls}
	\int\limits_{A^\mathrm{min,e}(\alpha_2)}^{A^\mathrm{max,e}(\alpha_2)} \left[\tilde F_1(z^\mathrm{e}(\alpha_2)) - \tilde F_2(z^\mathrm{e}(\alpha_2); k_2)\right]^2 dz,
\end{equation}
which in most regimes of the tip motion has only one global minimum lying in the deep valley defined by the curve $\alpha_2^\mathrm{true}k_2^\mathrm{true}$. Moreover, increasing the reconstructed polynomial power, $P_z$, makes this valley deeper and hence more resistant to noise [see also Ref.~\onlinecite{forchheimer}]. This method allows estimation of the product $\alpha_2k_2$ with higher accuracy than $\alpha_2$ and $k_2$ separately. 

\begin{figure}
\includegraphics[height=40mm,angle=0]{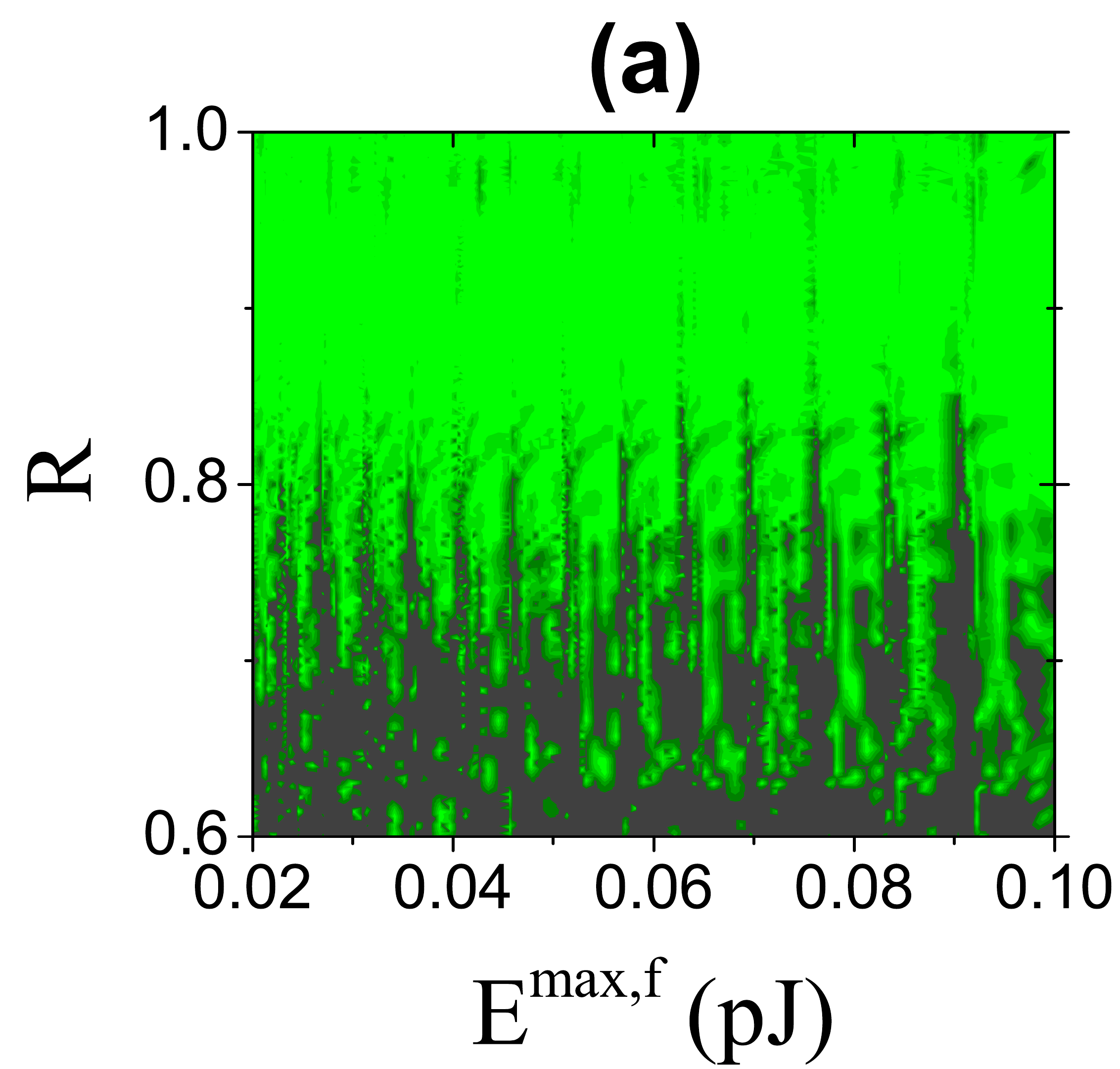}
\includegraphics[height=40mm,angle=0]{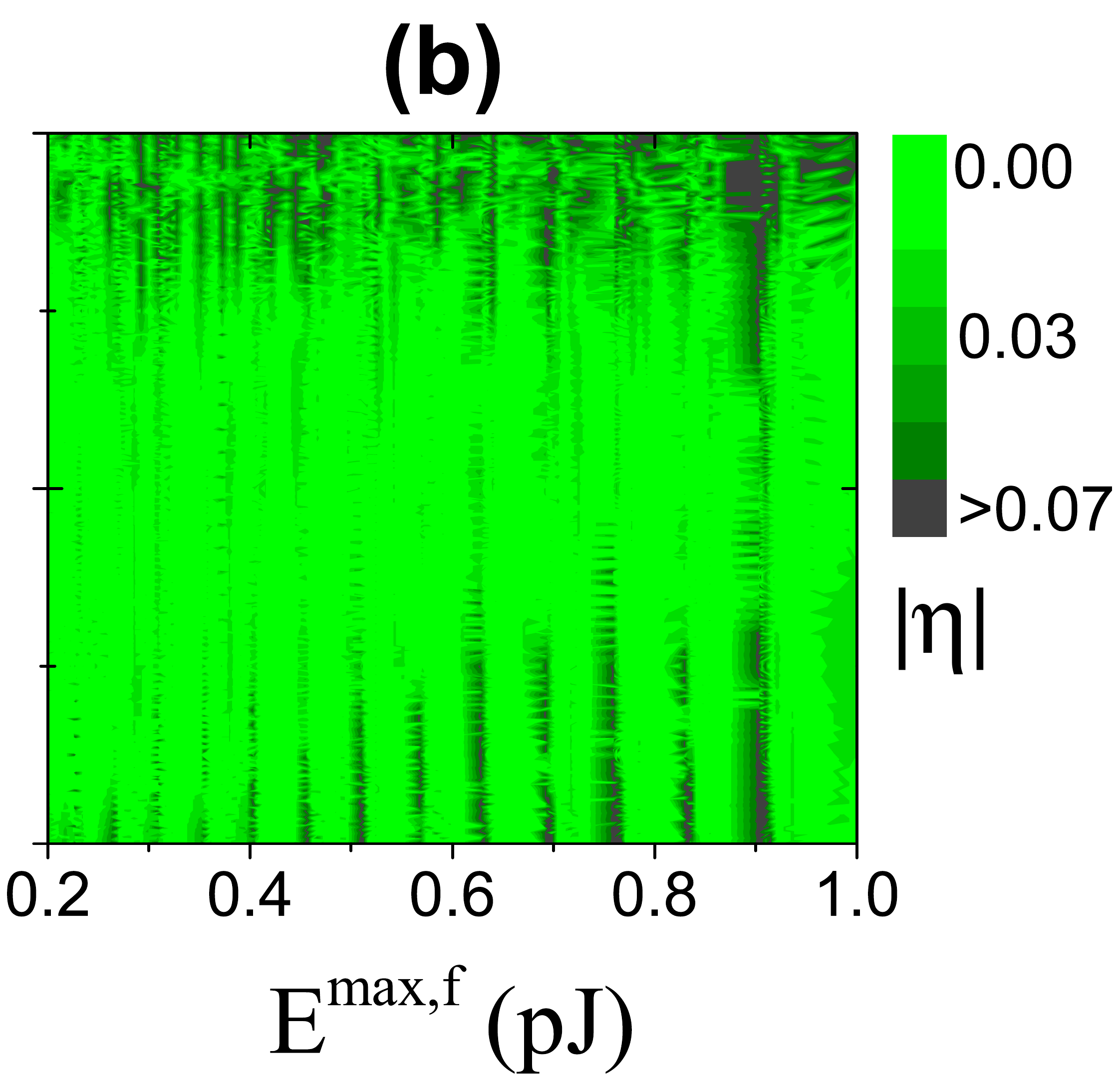}
 \caption{(Color online) Absolute value of the relative calibration error $\eta$ of $k_2\alpha_2$ as a function of the ratio $R=h/A^{\mathrm{max,f}}$ and total maximum free oscillation energy $E^\mathrm{max,f}$ for the (a) soft and (b) stiff cantilever.}
\label{fig:nonlinear_errors}
\end{figure}

Figure \ref{fig:nonlinear_errors} shows the absolute value of the relative error $\eta = 1 - k_2\alpha_2 / k_2^\mathrm{true}\alpha_2^\mathrm{true}$ plotted in the plane of maximum free oscillation energy $E^\mathrm{max,f} = (k_1(A_1^\mathrm{max,f})^2 + k_2(A_2^\mathrm{max,f})^2) / 2$ and the ratio $R=h/A^{\mathrm{max,f}}$. The relative calibration error is small over a wide range of oscillation energy and probe height. However, the vertical periodic stripes of lower error correspond to a large value of the ratio $A_1^\mathrm{max,f}/A_2^\mathrm{max,f}$. Experimentally, one can check the stability of calibration by comparing different probe heights and oscillation energies. Finally, the stiff cantilever has a wider region of low error because higher oscillation energy effectively weakens the nonlinearity.

%-------------------------------------------------------------------------------
{\it Calibration using a linear tip-surface force.---}
When the interval of the engaged tip motion is small, the tip-surface force [Eq.~(\ref{eq:force_model})] can be linearized. In this case, it is possible to obtain the explicit expression for the stiffness using a linear model $\tilde F$ with one unknown parameter\cite{Note3}
\begin{equation}
\label{eq:stiffness_linear}
	k_2=\left|\frac{\hat G_1(\omega)}{\hat G_2(\omega)}\right| \left|\frac{\Delta\hat V_1(\omega)}{\Delta\hat V_2(\omega)}\right|k_1 .
\end{equation}
As previously mentioned, the multimodal drive at the resonant frequencies \cite{Note4} $\omega_1$ and $\omega_2$ produces enough response components in order to find $k_2$. The corresponding domain of the engaged tip motion and eigenmode sensitivity to the force are defined by the energy scale factor $k_n^{-1}\hat G_n(\omega_n)$, therefore calibration of the softer cantilever can be performed with higher accuracy. While for the stiff cantilever, really small drive amplitudes are required for acceptable calibration results. Near the surface, the force is highly nonlinear making the tip prone to sudden jumps to the contact. From an experimental point of view, probing only the attractive part of the interaction with small oscillation amplitudes protects tip from possible damage since the dissipation is almost zero in this regime.
\begin{figure}
\centering
\includegraphics[width=40mm,angle=0]{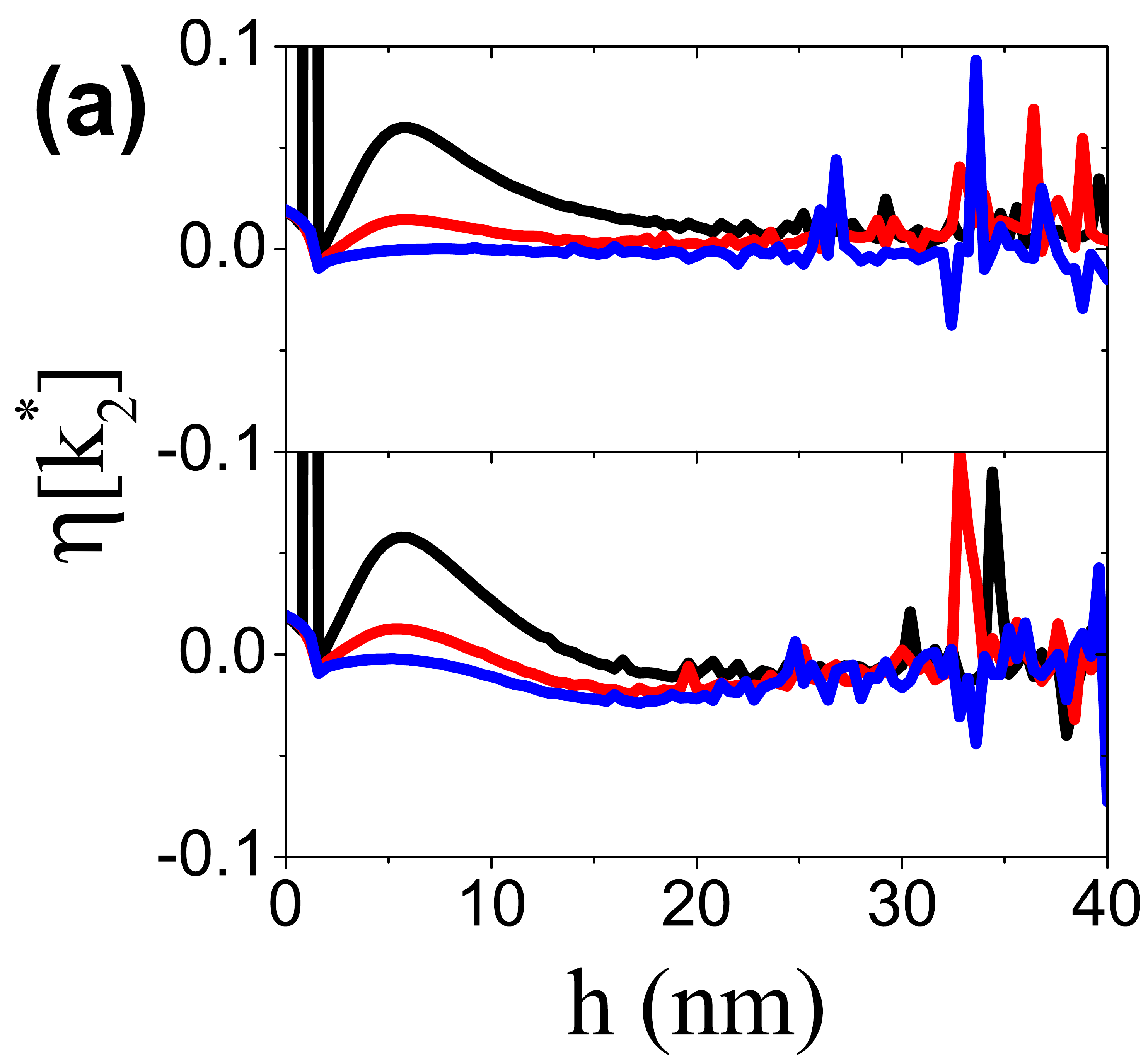}
\includegraphics[width=40mm,angle=0]{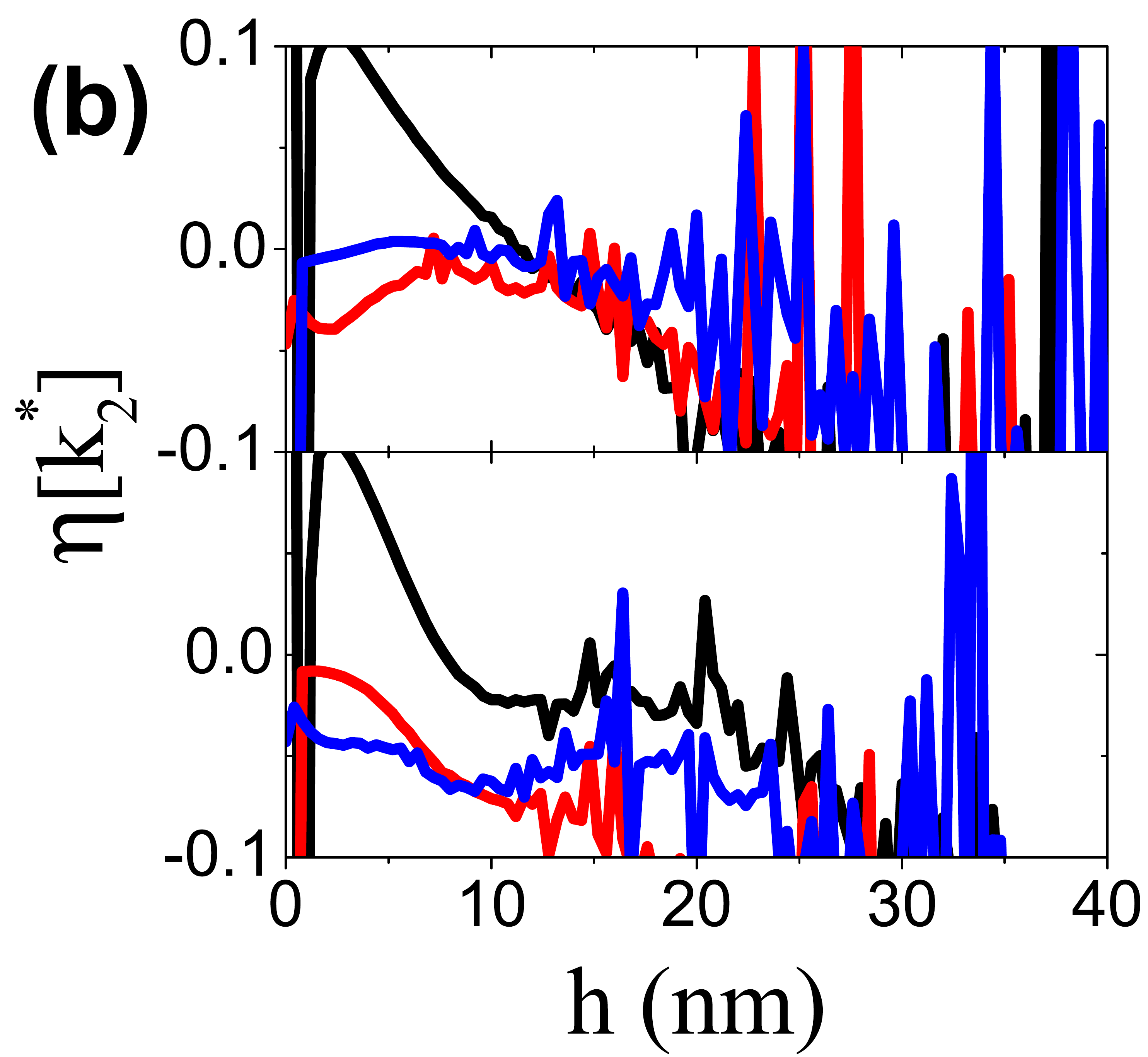}
 \caption{(Color online) Relative calibration error of the calibrated stiffness $k_2$, $\eta=1-k_2/k_2^\mathrm{true}$, using Eq.~(\ref{eq:stiffness_linear}) for two different cantilevers: (a) soft and (b) stiff, with different free eigenmode oscillations amplitudes: $A_1^\mathrm{f}=1$~nm (top), $A_1^\mathrm{f}=3$~nm (bottom), $A_2^\mathrm{f}=$ 0.1~nm (blue), 1.0~nm (red) and 2.0~nm (black).}
\label{fig:linear_errors}
\end{figure}
Finally, the linear method is dependent on the unknown higher eigenmode amplitude, $A_n$. Since it is not known {\it a priori}, one can use the following formula to try to make a rough guess given the known amplitude of the first mode 
\begin{equation}
\label{eq:A2_esteem}
	A_n^\mathrm{f} = \left|\frac{\hat G^\mathrm{piezo}(\omega_n)}{\hat G^\mathrm{piezo}(\omega_1)}\right| \frac{k_n}{k_1} \frac{Q_1}{Q_n} A_1^\mathrm{f},
\end{equation}
where $\hat G^\mathrm{piezo}$ is a transfer function of the piezoelectric shaker.

%-------------------------------------------------------------------------------
{\it Summary.}---We outlined a theoretical framework for experimental calibration of cantilever parameters using the tip-surface force with one-point measurement using a multimodal drive. The proposed approach does not require any knowledge of the cantilever's geometry or the tip-surface interaction form. In the tapping mode, the method possesses high calibration accuracy independently of {\it a priori} unknown amplitude of the higher eigenmode. Calibration in the non-contact attractive mode with small oscillation amplitudes keeps the tip maximally pristine.

%-------------------------------------------------------------------------------
\begin{acknowledgments}
This work is supported by KTH, Nordita, DOE, VR VCB 621-2012-2983 and the Knut and Allice Wallenberg Foundation.
\end{acknowledgments}

%\bibliography{calibration}


\end{document}